\documentclass[12pt]{article}

\newcommand{\Reals}[1]{\mathbb{R}^{#1}}

\newcommand{\A}{\mathcal{A}} 
 
\newcommand{\B}{\mathcal{B}} 

\newcommand{\X}{\mathcal{X}} 
\newcommand{\x}{x} 

\newcommand{\Y}{\mathcal{Y}}

\newcommand{\binaryset}{\{0,1\}}

\newcommand{\reals}{\mathbb{R}}

\newcommand{\simp}{\mathcal{O}}

\newcommand{\Policy}{\mathcal{F}}

\newcommand{\Expect}[1]{\mathop{\mathbb{E}_{#1}}}

\newcommand{\norm}[1]{\left\lVert#1\right\rVert}

\newcommand{\Wass}{\mathcal{W}}
\newcommand{\optcpl}{\pi}

\newcommand{\protecc}{\mathcal{S}}
\newcommand{\prot}{s}
\usepackage{dsfont}
\usepackage{microtype}
\usepackage{subfiles} 
\usepackage{graphicx}
\usepackage{subfigure}
\usepackage{amsmath}
\usepackage{amsthm}
\usepackage{mathtools}
\usepackage[margin=1in]{geometry}
\usepackage{amsfonts}
\usepackage{booktabs} 
\usepackage{hyperref}
\usepackage{authblk}
\usepackage{float}
\usepackage{natbib}

\theoremstyle{definition}
\newtheorem{definition}{Definition}[section]
\newtheorem{remark}{Remark}[section]
\newtheorem{example}{Example}[section]

\author[1]{Kweku Kwegyir-Aggrey}
\author[2]{Rebecca Santorella}
\author[3]{Sarah M. Brown}

\affil[1]{Department of Computer Science, Brown University}
\affil[2]{Division of Applied Mathematics, Brown University}
\affil[3]{Department of Computer Science, University of Rhode Island}
\date{\vspace{-1.6cm}}

\title{Everything is Relative: Understanding Fairness with Optimal Transport}

\begin{document}

\maketitle

\begin{abstract}
To study discrimination in automated decision-making systems, scholars have proposed several definitions of fairness, each expressing a different fair ideal. These definitions require practitioners to make complex decisions regarding which notion to employ and are often difficult to use in practice since they make a binary judgement a system is fair or unfair instead of explaining the structure of the detected unfairness. We present an optimal transport-based approach to fairness that offers an interpretable and quantifiable exploration of bias and its structure by comparing a pair of outcomes  to one another. In this work, we use the optimal transport map to examine individual, subgroup, and group fairness. Our framework is able to recover well known examples of algorithmic discrimination, detect unfairness when other metrics fail, and explore recourse opportunities.

\end{abstract}

\section{Introduction}
Machine learning-based decisions in sensitive domains such as criminal justice \cite{predictandServe} or banking \cite{bankingExample} have come under significant scrutiny due to their ability to amplify, re-invent, and create patterns of discriminatory behavior against already marginalized and vulnerable groups. Many scholars have attempted to prevent algorithmic discrimination by engineering \textit{fair} algorithmic solutions. These efforts \cite{zafar2017fairness, hardtEqualOppurtunity, woodworth2017learning} rely on the assumption  that if it can be proven or demonstrated that a classifier exhibits \textit{fair} behavior with respect to some agreeable notion of fairness, this classifier will no longer produce discriminatory outcomes. In contrast, critical scholars instead assert that discrimination by algorithmic systems is a symptom of broader systematic and historic forms of oppression \cite{selbst2019fairness}, which cannot be ameliorated through fair algorithmic solutionism.  We suggest that the key tension between these viewpoints relies on resolving the following question: \textit{If machine learning can be used to examine the discrimination present in algorithmic solutions, how can bias\footnote{In this framing, we view bias as measurement of discrimination relative to some fair ideal. } be meaningfully quantified and studied within this context to prevent further discriminatory outcomes?}

Despite countless examples of bias in automated decision making, a precise definition of fairness in this context remains elusive.  Philosophical and legal doctrine have proposed several conceptualizations of fairness such as disparate impact and equal opportunity, but no single fair notion has become standard or ubiquitous.  Instead, each fairness notion exists to describe idealized fair behavior in a different context or application.  The machine learning community approaches fairness similarly, usually by translating some abstract notion into a mathematical expression and then declaring classifiers as fair if they satisfy said expression. We consider two types of fair criteria: individual fairness \cite{dwork2012fairness} and group fairness.  The first postulates that similar individuals should be treated similarly, regardless of protected group membership such as race or gender, while the second considers the context of these protected attributes and looks to achieve some notion of parity between groups.  For example, the group fairness notion of demographic parity suggests that fairness is achieved when a classifier's predictions are independent of individuals' protected attributes.

Although these fair definitions are tractable and succinct, they often lack sufficient nuance to examine fairness in the context of real sociotechnical systems \cite{hoffmanWhereFairnessFails}. For example, it has been shown that binary assumptions about protected group membership limit the applicability of fair algorithms \cite{hanna2020towards}.  While these assumptions make some fairness problems tractable, \cite{kearnssubgroup} demonstrated that they are not sufficient to compute fair outcomes over more complex groups such as a set of protected groups that may intersect \footnote{Our intersectional understanding of identity is based in the notion of Intersectionality Theory \cite{crenshaw2017intersectionality}}.  Moreover, these fairness definitions can suggest that a system is fair or unfair but are unable to provide actionable recourse as to how to amend the bias, further rendering these definitions deficient in practice \cite{corbettdavies2018measure}.  


Our solution is to evoke an understanding of fairness that does not rely on closed-form definitions. Instead, we propose that a set of outcomes can only be deemed unfair if it deviates strongly from a set of outcomes that are deemed fair. In other words, fairness is a property that can be evaluated by comparing sets of outcomes. This intuition can be made tractable by viewing sets of outcomes as probability distributions and then leveraging some statistical difference measure to quantify the differences between outcomes. 

\begin{example}
\label{aa}
A university admissions process may be considered unfair if it admits cohorts that are not sufficiently representative or diverse.
This complaint follows from comparing the admitted cohort to another, possibly hypothetical, \textit{fair} cohort.
This notion of fairness does not necessarily require parity in classification - not all student groups must be admitted with the same frequency - which differentiates this intuition from current fair algorithmic practices.  
\end{example}
We advocate for the Wasserstein distance as the best statistical distance metric to compare outcome distributions. The Wasserstein distance has many advantages over other distribution metrics such as KL-Divergence or Total Variation distance (we refer the reader to \citet{wassersteinGAN} for a summary of these advantages); however, there are two advantages we would like to highlight.  First, the Wasserstein distance can be computed between distributions that are not defined over the same supports.  This flexibility means that we can compare unfairness across two seemingly disparate groups of observations. Additionally, computing the Wasserstein distance yields a probabilistic coupling matrix, which describes how to transport one set of outcomes onto another. We can leverage this transport map to better understand the structure of observed biases and offer actionable suggestions to reduce bias in automated systems. 

\subsection{Contributions}
To circumvent the limitations of the definition-based fairness approach, we provide an optimal transport-based framework for exploring and quantifying bias in automated systems. We compute the Wasserstein distance between outcomes for different groups to understand how one set of outcomes may be transformed into another. This distance is able to detect bias when other fairness metrics fail, and we can leverage the transport map used in the distance calculation to quantify individual, subgroup, and group bias. This approach allows us to recover biases previously found in the COMPAS and German credit datasets, as well as offer possible actions for recourse.



\section{Background}
\label{background}

In this section, we introduce key definitions in fairness and relevant mathematical background on optimal transport. 
\subsection{Preliminaries}
Allow $\X \subseteq \Reals{k}$ to be some set of features and $\Y = \binaryset$ to be a binary set of possible outcomes for some decision. We assume that $\protecc = \binaryset$ describes the protected group membership of an individual. A classifier is a function $h: \X \rightarrow \binaryset$ which outputs a classification decision $h(\x)$ for each individual $\x \in \X$.  We assume there exists a distribution $\mathcal{D}$ over the set $\mathcal{Z} = \X \times \protecc \times \Y$, which takes on values  $(x, \prot, y)\sim\mathcal{D}$. 

\subsection{Parity-Based Fairness}
Below we review common definitions of group fairness that we will use as points of comparison. 

\begin{definition}[Demographic Parity]
A classifier is fair under demographic parity if the classifier's predictions are independent from the sensitive attribute \cite{demographicParityCalders}:
\begin{align}
    \Expect{\mathcal{D}}[h(X)|\protecc=1] = \Expect{\mathcal{D}}[h(X)|\protecc=0].
\end{align}
\end{definition}


\citet{chouldechova2016fair} demonstrated that satisfying demographic parity requires each protected group to have the same distribution of features.  Most data distributions violate this condition, and in that case, enforcing demographic parity can have an adverse effect on classifier performance  \cite{zafar2017fairness, tooRelaxed, HuWelfare2020}. To avoid this performance loss, difference of demographic parity (DDP) \cite{DDP} relaxes the strict independence assumption to,
\begin{align*}
    DDP(f) \coloneqq \Expect{\mathcal{D}}[h(X)|\protecc=1] - \Expect{\mathcal{D}}[h(X)|\protecc=0],
\end{align*}
and fairness is enforced by minimizing this difference. 

Although DPP can increase the number of tractable outcomes, it suffers from many of the same pitfalls as demographic parity.  Namely, it relies on binary protected groups, offers no quantification of unfairness \footnote{In Section~\ref{disparate_impact} we discuss disparate impact, which quantifies a lack of demographic parity}, and does not provide additional information regarding the structure of observed unfairness.   

Equal opportunity \cite{hardtEqualOppurtunity} achieves fairness by equalizing the per-group true positive rates of a classifier.  
\begin{definition}[Equal Opportunity] A classifier is fair under equal opportunity if its predictions have the same true positive rates across sensitive attribute membership:
\begin{align*}
\Expect{\mathcal{D}}[h(X)|Y=1, \protecc=1] = \Expect{\mathcal{D}}[h(X)|Y=1, \protecc=0].
\end{align*}
\end{definition}
Although equal opportunity differs from demographic parity in that it can be enforced despite difference in group conditioned feature distributions, the two expressions share many of the same limitations such as the inability to examine the structure of unfairness.  We refer the reader to \cite{corbettdavies2018measure} for a more complete analysis of parity-based fairness definitions. 

\subsection{Individual Fairness}
Individual fairness \cite{dwork2012fairness} is summarized by the intuition that ``similar individuals should be treated similarly.'' Although this definition is semantically appealing, choosing a reasonable similarity metric is often task-specific, thus rendering this notion of fairness untenable across domains \cite{rothCritique}. Furthermore, there are many instances, such as Example~\ref{aa}, where fairness may require treating dissimilar individuals similarly, which is not captured by this definition. 


\subsection{Optimal Transport}

To measure the difference between sets of outcomes, we use optimal transport, which seeks the most cost-effective way to transform one probability distribution into another \cite{peyre2019computational}. If we think of distributions as probabilities on point masses, then optimal transport finds the minimal effort way to move masses from one domain onto masses in another domain. The Kantorovich formulation allows the transport map to split masses between domains, meaning that the map does not need to be a bijection.

Formally, given two measures $\mu_1$ and $\mu_2$ defined on the measure spaces $\A$ and $\B$, respectively, the optimal transport problem seeks an optimal coupling $\optcpl$ from the set:
\begin{multline}
    \Pi(\mu_1, \mu_2) = \{ \optcpl \in P(\A \times \B) : \optcpl(A \times \B) = \mu_1(A)  \text{ for }  A \subset \A, \\ \optcpl(\A \times B) = \mu_2(B) \text{ for } B \subset \B \}. 
\end{multline}
The coupling $\optcpl$ is a new probability distribution over the product space $\A \times \B$ that assigns probabilities to relate masses from domain $\A$ to masses in domain $\B$. The constraints ensure that the marginal distributions of $\optcpl$ match the original input distributions  $\mu_1$ and $\mu_2$. The full optimal transport problem with constraints is given by,
\begin{equation}
    \min_{\optcpl \in \Pi(\mu_1, \mu_2)} \int_{\A \times \B} c(a, b)  d \optcpl(a, b),
    \label{ot}
\end{equation}
where $c: \A \times \B \to \mathbb{R}$ is a cost function that measures how difficult it is to move mass from domain $\A$ to $\B$. 

The $p^\text{th}$ Wasserstein distance is a special case of this optimal transport problem on the metric measure space $(\A, d)$ with set $\A$ and distance $d: \A \times A \to \reals$ when the cost function is $c(a,b) = d(a,b)^p$: 
\begin{equation}
      W_{p}(\mu_1, \mu_2) \equiv \left(\inf_{\optcpl\in \Pi(\mu_1,\mu_2)} \int_{\A \times \A} d(a_1, a_2)^{p} d\optcpl(a_1,a_2)\right)^{\frac{1}{p}}.
      \label{Wasserstein}
\end{equation}
Wasserstein distance is commonly used in machine learning to measure distances between probability distributions. 

Given data $A = \{a_{i}\}_{i=1}^{n_1}$ from $\A$ and $B= \{b_{j}\}_{j=1}^{n_2}$ from $\B$, the marginal distributions can be expressed as 
\begin{equation}
    m_1 = \sum_{i=1}^{n_1} p_i \delta_{a_i} \text{ and } m_2 = \sum_{j=1}^{n_2} q_j \delta_{b_j},
\end{equation}
where $\delta_{a}$ is the Dirac measure, and the set of admissible couplings are the matrices 
\begin{equation}
    \Pi(m_1,m_2) = \{ \optcpl \in \mathbb{R}_+^{n_1 \times n_2} : \optcpl \mathds{1}_{n_2} = m_1, \ \optcpl^T \mathds{1}_{n_1} = m_2 \}. 
\end{equation}
Intuitively, each entry $\optcpl(a_i,b_j)$ in a coupling matrix says how much mass from point $a_i$ in $A$ should be mapped onto point $b_j$ in $B$.  Similarly, the cost can be described as a matrix $C \in \mathbb{R}^{n_1 \times n_2} $, where $C_{ij} = c(a_i, b_j)$ captures the cost of moving point $a_i$ onto point $b_j$. Using this setup, the discrete optimal transport problem is
\begin{equation}
    \min_{\optcpl \in \Pi(m_1,m_2)} \langle \optcpl, C \rangle
    \label{discreteOT},
\end{equation}
which can be solved with minimum cost flow solvers \cite{bonneel2011displacement}.

\section{Method}
\label{method}
 
In this section, we introduce our optimal transport-based framework and explain how we quantify bias in individuals, subgroups, and groups. 
 
\subsection{Comparative Fairness}
We are interested in comparing the differences between two decision-making policies with respect to a given set of outcomes, where outcomes are probability vectors over possible predicted labels. To capture more nuance in algorithmic decisions, we extend the concept of a classifier to a policy, which expresses outcomes (or predictions) as probabilities rather than binary labels. 

\begin{definition}[Policy]
A policy $\Policy: \X \rightarrow \simp(\Y)$ is a mapping  from the feature space to the space of outcomes. 
\end{definition}


Denote the distribution on outcomes generated by observing a policy $F$ on the space $\X$ by $\mu_\Policy^\X$. Given policies $\Policy_{1}$ and $\Policy_{2}$ on subsets $\A$ and $\B$ of $\X$, we use the Wasserstein distance to measure the differences in outcomes they produce:
\begin{align}
    \Wass_2(\mu_{\Policy_{1}}^\A, \mu_{\Policy_{2}}^\mathcal{B}).
\end{align}
This parametrization provides control over what comparison the distance captures. By setting $\A$ and $\B$ to be the same population, we measure the impact of different policies. Alternatively, we can take $\Policy_1$ and $\Policy_2$  to be the same policy to study its effects on different populations. 

\begin{figure}[t]
\centering
\includegraphics[width=0.48\textwidth]{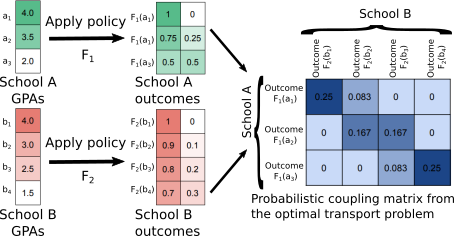}
\caption{Schematic of optimal transport applied to the school admissions example \ref{disparate_impact}. Policies are applied to the two populations to generate a transport map between their outcomes.}
\label{fig:method}
\end{figure}

In practice\footnote{We implement this method with the Python Optimal Transport toolbox (\url{https://pot.readthedocs.io/en/stable/})}, we do not have access to the true distributions of the outcomes, so, we approximate the outcome distributions with uniform distributions as is standard in computational optimal transport \cite{peyre2019computational} and compute $W_2(m_{\Policy_1}^A, m_{\Policy_2}^B)$:
\begin{equation}
    m_{\Policy_1}^A = \frac{1}{n_1} \sum_{i=1}^{n_1} \delta_{\Policy_1(a_i)} \text{ and }
    m_{\Policy_2}^B = \frac{1}{n_2} \sum_{j=1}^{n_2} \delta_{\Policy_2(b_j)}. \\
\end{equation}
In the above empirical distributions, the outcomes are generated by applying the policies to data $A = \{a_{i}\}_{i=1}^{n_1} \sim \A$ and $B= \{b_{j}\}_{j=1}^{n_2} \sim \mathcal{B}$. In this case, the cost matrix is computed by:
\begin{equation}
    C_{ij} = \norm{ \Policy_{1}(a_{i}) -  \Policy_{2}(b_{i}) }_2. 
\end{equation}
Figure~\ref{fig:method} gives an overview of this process. 

This computation reveals the magnitude of difference between the outcomes from policies $\Policy_1$ and $\Policy_2$.  Allow $\optcpl$ to be the optimal coupling associated with the above Wasserstein distance. This coupling is a transport plan describing how to split the mass of outcomes from $\Policy_{1}$ onto outcomes from $\Policy_{2}$ to produce similar behavior (or vice-versa). In particular, the entries of the normalized row $n_1 \optcpl_i$ describe how to split the mass of outcome $\Policy_1(a_i)$ onto the outcomes $\{ F_2(b_{j}) \}_{j=1}^{n_2} $. Since the outcome vectors result from applying policies to individuals in the datasets, we can also trace back the policy mappings and interpret the transport map between individuals in the two sets. 

\subsection{Measuring Individual and Group Bias}

While the Wasserstein distance measures how different two sets of outcomes are, it does not take into account differences between the underlying populations to which each policy is applied. Thus, to compare two policies and examine their biases, we need to introduce metrics which take the feature space into account. Each outcome $\Policy_1(a)$ corresponds to some individual $a$, so we can think of the coupling as mapping individuals in $A$ to individuals in $B$ through their outcomes. 

Allow $\Gamma(a,\optcpl)$ to be the set of all individuals that $a \in \A$ is mapped to under the coupling $\optcpl$:
\begin{align}
    \Gamma(a, \optcpl) = \{b \in B : \optcpl(a,b) > 0\}.
\end{align}
Suppose we are able to quantify how different individuals are with some distance on the feature space $d: \mathcal{X} \times \mathcal{X} \to \mathbb{R}$. Then, using this set we can define a notion of comparative individual bias.

\begin{definition}[Individual Bias]
\label{individual_bias}
The bias an individual $a$ experiences across policies is measured by the expectation:
\begin{align}
     u_{\X}(a) = \Expect{b \sim \optcpl(a, \cdot)}[d(a,b)].
     \label{eq:individual_bias}
\end{align}

\end{definition}

We can extend this definition of individual bias to subgroups and groups. Consider a partitioning of the input space $\X$ into discrete groups $\mathcal{G} = (G_{1}, G_{2},...G_{n})$ where $G_{i} \subseteq \mathcal{X}$ such that $\X = \bigcup_{i=1}G_{i}$. We say that the bias experienced by some group is then a sum of the individual bias terms for members of that group.
\begin{definition}[Group Bias]
The bias (comparatively) experienced across policies for a group $G$ is measured by:
\begin{align}
    U_\X(G) = \sum_{g \in G} u_{\mathcal{X}}(g).
\end{align}
\end{definition}

\begin{remark}
We also can compare groups directly rather than across the whole population. If for all $i,j \leq |\mathcal{G}|$ all groups are disjoint $G_{i} \cap G_{j} = \emptyset$ then the group bias measurement can be additively decomposed as
\begin{align*}
    U_\X(G_{i}) = \sum_{j \leq |\mathcal{G}|} U_{G_{j}}(G_{i})
\end{align*}
where each $U_{G_{j}}(G_{i})$ is the amount of bias measured in group $G_{i}$ that is the result of comparison with individuals in group $G_{j}$.
\end{remark}

The success the above framework relies on the existence of a metric $d: \mathcal{X} \times \mathcal{X} \to \mathbb{R}$, which can suitably compute differences between individuals. While taking the Euclidean distance over feature space is often possible, we leave the choice in distance metric as a hyperparameter in our problem formulation to allow for different measures of bias. We leverage this in our exploration of recourse in \ref{sec:recourse}.
\section{Experimental Results}
In this section, we demonstrate the flexibility of our optimal transport framework. First, we show that the Wasserstein distance detects bias when disparate impact fails. Then, we show that we can use the coupling matrix to recover known biases in the COMPAS dataset. Finally, we use the optimal coupling to explore recourse options for individuals, subgroups, and groups in the German credit dataset.

\subsection{Wasserstein Distance Detects Bias when Disparate Impact Fails}
\label{disparate_impact}

To motivate the Wasserstein distance for fairness evaluation, consider the fictitious Blue College, who are looking to audit their admissions data for bias. Blue's applicants come from two secondary schools: expensive private School A and public School B.  Blue College would like to check that their admissions policy is not biased with respect to an applicant's secondary school. Looking at previous admissions data, Blue College observes the following pattern:
\begin{enumerate}
    \item For \textit{all} students in School A, the probability of being accepted to Blue College is $P(Y=1|X=x) = P(Y=1) = 0.25$. 
    \item For students in School B, the probability of being accepted is $P(Y=1|X=x) = 1$ for the students with GPAs in the top 25\% of GPAs in their class, and $P(Y=1|X=x) = 0$  for the bottom 75\%.
\end{enumerate}

Blue evaluates their decisions with the disparate impact criteria, which provides a measure of demographic parity. 
\theoremstyle{definition}
\begin{definition}[Disparate Impact (80\% Rule)]
A model is said to admit disparate impact if:
\begin{align}
    \frac{P(Y=1|\text{School}=B)}{P(Y=1|\text{School}=A)} \leq \tau = 0.8 .
\end{align}
\end{definition}

Blue college accepts $25\%$ of the applicants from both schools, which implies that $\frac{P(Y=1|\text{School}=A)}{P(Y=1|\text{School}=B)}$ = 1, and so the model does not admit disparate impact and is considered \textit{fair} with respect to this fairness definition.  More generally, if $P(Y=1|\text{School}=B) \le P(Y=1|\text{School}=A)$ then $\tau =1$ implies this is the \textit{fairest} possible outcome. Clearly, this model is not fair as it puts the bottom 75\% of students from School B at a disadvantage, while also offering an advantage to the top 25\% of students at school B.  

\begin{figure}[t]
    \centering
    \includegraphics[scale=.35]{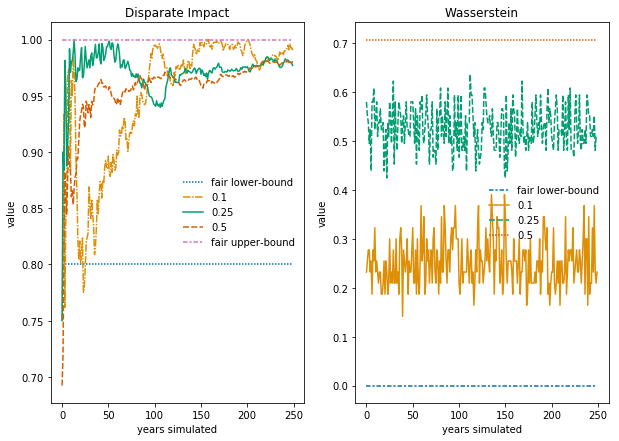}
    \caption{Under disparate impact, all rules converge to a fair policy despite structural inequity in the admissions policy. In contrast, the Wasserstein distance recovers the inequities at all rule levels.}
    \label{fig:di_stinks}
\end{figure}

We simulate several years of admissions according to this fictitious model with $\A$ and $\B$ as the students in the two schools, and $\Policy_1$, and $\Policy_2$ as the two described admissions patterns. In addition to the 25\% rule described above, we also evaluate a 10\% and 50\% rule.  A rule with a lower percentage indicates that fewer students from both schools are accepted by Blue College, and the $0\%$ rule indicates that no student from A or B is accepted. In Figure~\ref{fig:di_stinks}, we plot the Wasserstein distance and disparate impact score as aggregated over the years. 

\begin{figure*}[t]
    \centering
    \includegraphics[width=\textwidth]{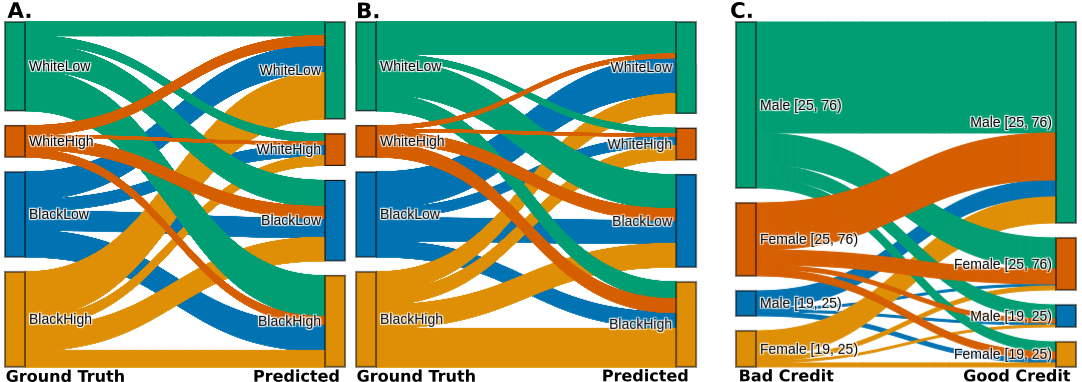}
    \caption{The transport map reveals how subgroups are mapped to one another from the ground truth outcomes to the predicted outcomes for
    \textbf{A.} COMPAS's predictions and \textbf{B.} equal opportunity classifer's predictions. For example, a large fraction of high-risk white individuals achieve classification outcomes from COMPAS similar to low-risk Black defendants when compared to ground-truth recidivism. These differences are less stark in the outcomes produced by the Equal Opportunity classifier.  \textbf{C.} The transport map recovers previously found biases in age and gender in the German credit dataset. }
    \label{fig:flows}
\end{figure*}

According to disparate impact, for any given year Blue College may appear fair or unfair, but the aggregate admissions data appears fair over time, despite obvious bias in the admissions process. Conversely, the Wasserstein distance reveals a fixed amount of bias over time and shows that the observed bias decreases as the rule tends towards zero. As the decision rule approaches the trivially fair scenario where all individuals from both schools are rejected, the Wasserstein distance decreases, indicating that the outcomes generated by the policies are moving closer together.

\subsection{Recovering Racial Bias in Recidivism Risk Scores}
\label{sec:compas}

We use the COMPAS \cite{COMPAS} dataset to show that the optimal coupling can be used to detect bias in subgroups. To audit the COMPAS tool for racial bias, Propublica \cite{larson2016we} collected  COMPAS scores, prior arrest records, and two year re-arrest records in Broward County, Florida. In the accompanying report, they found that the COMPAS tool assigned Black defendants risk scores that were disproportionately higher than those assigned to their white counterparts, even after controlling for criminal history.  As a result, Black defendants who did not recidivate were often misclassified as ``high-risk,"  whereas for white defendants, the opposite was true: those who \textit{did} get rearrested were often mistakenly assessed as ``low-risk." To corroborate these results, we apply our optimal transport framework to the COMPAS dataset.

We take $A = B$ to be the set of individuals in the COMPAS dataset and compute the Wasserstein distance between the observed (two years after initial arrest) and predicted (by COMPAS)  outcomes. For both the observed and predicted cases, we compute a logistic regression model, which takes as input relevant criminal history information
and actual and predicted recidivism labels as targets.  We use these logistic models to estimate the likelihood an individual recidivates under the two scenarios, denoted $\mu_{\Policy_{1}}^A$ and $\mu_{\Policy_{2}}^A$, respectively, where $\Policy_{1}$ denotes the true and $\Policy_{2}$ the predicted recidivism policies. By computing $W_{2}(\mu_{\Policy_{1}}^A, \mu_{\Policy_{2}}^A)$ we obtain the optimal coupling, which we then use to observe the differences between policies. 

\begin{figure*}[t]
    \centering
    \includegraphics[width=\textwidth]{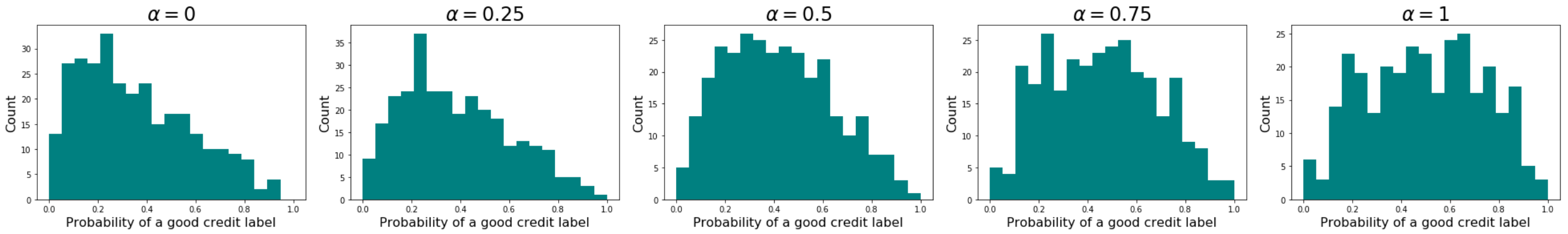}
    \caption{As $\alpha$ increases, increasing the weight of the actionable features of those with good credit in the interpolation (Equation~\ref{eq:interp}), the distribution of the probability of receiving a good credit label for those who had previously been labeled as bad credit shifts from being skewed towards low probabilities to a more uniform distribution. }
    \label{fig:alpha}
\end{figure*}

Our framework recovers the bias documented in the Propublica study.  First, we observe that as a result of the the discriminatory policy, the group bias score for Black defendants who are considered high-risk is  $5.88\%$ higher than the Black defendants who were considered low-risk.  For white defendants, however, we observe a $4.92\%$ increase in bias in the opposite direction; the bias experienced by white individuals who are considered low-risk is higher than that of white individuals considered high-risk.  This finding suggests that Black individuals who are considered high-risk and white individuals who are considered low-risk are more likely to attain similar predicted outcomes to an individual who is dissimilar to them in feature space, which is consistent with Propublica's observations.  Black individuals who are considered high-risk often have similar criminal histories to white individuals who were considered low-risk and vice versa. In Figure~\ref{fig:flows}A we show the group-wise decomposition of bias: 43.8\% of the outcomes attained by low-risk white individuals are similar to  outcomes predicted for high-risk Black individuals and 48.3\% of high-risk Black individuals are mapped to low-risk white individuals. 

Next, we train a third logistic regression model on the COMPAS labels.  This regression differs from the earlier two by employing an equal opportunity constraint based on \cite{zafar2017fairness} to make the predicted outcomes more fair. The fair classifier increases the true positive rates at test time for both the Black and white groups by ~10\%. Furthermore, in Figure~\ref{fig:flows}B, we can see that more of the subgroups are mapped to their appropriate counterparts. 

For the positive-predicted group, which we originally documented as discriminatory against mostly Black defendants, we see a decrease in unfairness from the original to the equal opportunity outcome distribution.  Moreover, when we decompose the source of remaining bias,
a larger share of individuals are mapped to defendants in their same predicted risk class.  This observed decrease in unfairness suggests that the fair classifier produces outcome distributions that are less dependent on race than the original COMPAS predictions.

\subsection{Actionable Recourse in the German credit dataset}
\label{sec:recourse}
To explore recourse opportunities with optimal transport, we use the German credit dataset \cite{German}, which contains loan applications at a bank and their classification as good or bad credit risks. It is commonly used for algorithmic recourse investigations \cite{gupta2019equalizing, karimi2020algorithmic, chen2020strategic, rawal2020interpretable}. We train a logistic regression classifier $\Policy$ on the features without the sensitive attributes (age and gender) to produce outcome probabilities for each individual and then compute the transport map between those who were classified as bad creditors, $A$, and those who were classified as good, $B$, under policy $\Policy$. Figure~\ref{fig:flows}C shows that this map recovers age and gender discrimination that has previously been found, namely that the policy favors older and male applicants. In particular 69.85\% of men aged $[18,25)$, 66\% of women aged $[18,25)$, and 63.21\% of women aged $[25,76)$ labeled as bad credit are disproportionately mapped to men aged $[25,76)$ labeled as good credit. 

\begin{figure*}[t]
    \centering
    \includegraphics[width = \textwidth]{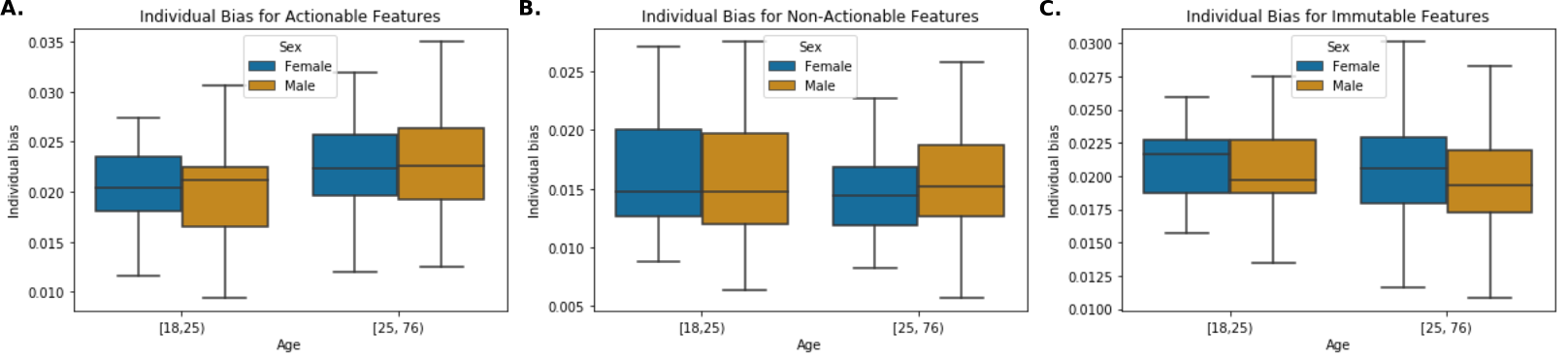}
    \caption{Individual bias in those labeled as bad creditors broken down by age and sex in the German credit dataset \cite{German} when distances are taken with respect to \textbf{A.} actionable, \textbf{B.} non-actionable, and \textbf{C.} immutable features. }
    \label{fig:german_bias}
\end{figure*}

Algorithmic recourse asks what actionable changes an individual can take to change their classification \cite{karimi2020survey}. To understand recourse options for individuals, we break up the features into three types: actionable (and mutable),  non-actionable (but mutable), and immutable (and non-actionable) (see Table~\ref{tab:german_credit} in the appendix for the full breakdown). For each type, we can compute the individual bias (Equation~\ref{eq:individual_bias}) while restricting the distance computation to features of that type. In Figure~\ref{fig:german_bias}, we plot the distribution of the individual biases for each type of feature when split by age and sex for those in the bad credit class.

We see that men aged $[25,76)$ with a bad credit classification have the highest actionable bias, suggesting that this group's actionable features are the furthest from those with a good credit classification. Similarly, women aged $[18,25)$ with a bad credit classification have the lowest actionable bias even though they receive far fewer good credit ratings. Since the actionable features mainly measure attributes that a \textit{fair} classifier should consider the most, credit history for example, this difference reinforces the claim that the policy discriminates along age and gender. For both non-actionable and immutable features, the differences in individual bias are stronger across sex than across age. This result suggests that this classification algorithm disadvantages women for attributes that they cannot change. 

For each individual $a$ classified as having bad credit, we can compute $\Gamma(a, \optcpl)$, the group of people with good credit that they were mapped to under the optimal coupling $\optcpl$. The product $ n_1 \sum_{b \in \Gamma(a, \optcpl)} \pi(a,b) b $ projects individual $a$ onto their counterparts in group $B$. Using this product, we can interpolate an individual's actionable features with those they were mapped to with weight $\alpha$ through
\begin{equation}
    (1-\alpha) a_{\text{actionable}} + \alpha n_1 \sum_{b \in \Gamma(a, \optcpl)} \pi(a,b) b_{\text{actionable}}. 
    \label{eq:interp}
\end{equation}
We can then reapply policy $\Policy$ to an individual from the bad credit group with their original immutable and non-actionable features and interpolated actionable features to see how their classification changes in response to these actions. Figure~\ref{fig:alpha} shows how the probability of being classified as having good credit increases with $\alpha$, implying that making actionable changes to these features can affect outcomes. As $\alpha$ increases, the  distributions shift from centering around low probabilities of receiving a good credit label to more uniform probabilities of receiving a good credit label, implying that changing these actionable features will improve an applicant's odds of success. 

We can also examine changes in the actionable features of those who were reclassified to understand which features were the most important (see Figure~\ref{fig:german_changes} in the appendix). Previous studies tend to focus on recourse accuracy (how many people were successfully reclassified as a result of actions taken) rather than which features were deemed most important, so we cannot compare our results directly.  Our results suggest that having a critical account or credit at other banks and not having a guarantor or co-applicant most positively affect classification and having no credits or credit repaid duly most negatively affect classification. Having a savings account has a large impact on classification; no savings account or more than $100$ DM in a savings account positively impacts classification, and having less than $100$ DM negatively affects classification.

\section{Related Work}

Our fairness philosophy builds on work by \citet{kearnssubgroup} and \cite{friedler2016impossibility}. \citet{kearnssubgroup} seek to reduce the divide between group and individual fairness by considering statistical notions defined over a large number of subgroups.  We also embrace a view of fairness beyond protected group attributes; however, our work differs in that it provides an interpretable framework to study bias as well as allowing notions of fairness to be flexibly instantiated. We also draw inspiration from \citet{friedler2016impossibility}, who argue that algorithmic fairness must consider the relationship between outcome and feature spaces. This philosophy inspired our analysis of fairness in the outcome space rather than in the feature space. 

Individual fairness \cite{dwork2012fairness} is often criticized due to its reliance on the existence of similarity metrics on features and outcomes \cite{rothCritique}.  While our definition of individual bias also relies on a distance between features, we are able to examine bias directly through the coupling matrix or vary the distance as a hyperparameter to fully explore possible relations in feature space. Thus, in our case, these distance parameters provide meaningful flexibility and insight.

Several other works \cite{silviaOptimal, zehlikeOptimal, gordaliza2019obtaining, wassersteinFairClassification, feldman2015certifying, johndrow2019algorithmLum, chiappa2021ContinuousOptimal} use optimal transport in the context of algorithmic fairness to compute a bias-neutral feature representation for downstream classification tasks using Wasserstein barycenter representations. Our work is not concerned with computing fair classifiers, but rather studying and auditing for sources of bias, which most closely relates to \cite{black2020fliptest}. 

Through FlipTest, \citet{black2020fliptest} use optimal transport to match individuals in different sensitive attribute groups and then test how changing an individual's group membership affects their classification. FlipTest uses the Monge formulation of optimal transport, which finds a permutation between the groups, rather than our more flexible Kantorovich formulation, which can split an individual's mass in the transport map. Additionally, their transport map is based on distributions on features, while ours is based on distributions on outcomes. The authors also use the set of individuals whose classification flipped to study which features were most important in the model. This approach is somewhat similar to our recourse experiments, but it audits the effects of all of the features, not only the actionable ones. Similarly, both \cite{taskesen2020statistical} and \cite{xue2020auditing} use optimal transport to measure changes in classification under perturbation to input features. 

\section{Discussion}
We have presented an optimal transport-based framework for auditing automated decision making that allows us to explore bias in individuals, subgroups, and groups. The flexibility in our framework allows us to test and compare multiple notions of fairness simultaneously; we can use the Wasserstein distance to quantify bias or leverage the coupling matrix to study the structure of bias. Our notion of comparative fairness means that we can decide what frame of reference to use to examine bias. Through the COMPAS example, we have demonstrated that optimal transport can investigate the impact of two different policies (predicted recidivism and ground truth recidivism) on the same population as well as the subgroups within that population. Furthermore, the transport map elucidates structures of bias and allows us to quantify it. In studying the German credit dataset, we have shown that optimal transport can audit automated decisions after they are made as well as offer potential recourse opportunities. 

One of the limitations of the standard optimal transport framework is that all of the mass from one distribution must be transported onto all of the mass of the other distribution. In the context of fairness, this limitation means that outliers and unevenly represented subgroups may not be mapped to their appropriate counterparts. To address this, we could modify our framework to use unbalanced or partial optimal transport. Additionally, computing transport maps between large datasets can be computationally intensive, so we could use entropically regularized optimal transport to audit larger datasets. Future work will focus on addressing these limitations as well as formalizing the  mathematical theory and its connections to previous definitions of fairness.

\clearpage
\newpage

\bibliographystyle{abbrvnat}
\bibliography{references/bib}

\clearpage
\newpage

\renewcommand{\thefigure}{A\arabic{figure}}
\renewcommand{\thetable}{A\arabic{table}}
\setcounter{section}{0}
\setcounter{figure}{0}

\section*{Appendix}

\subsection*{Computing Individual and Group Bias}

Our definitions of individual and group bias assume that we can measure the entire population $\X$. In practice, we have data $A = \{a_{i}\}_{i=1}^{n_1} \sim \A$ and $B= \{b_{j}\}_{j=1}^{n_2} \sim \mathcal{B}$, which requires slight changes for computations. 

\begin{definition}[Individual Bias]
The bias an individual $a$ from group $\A$ experiences in comparison to group $B$ across policies is measured by the expectation:
\begin{align}
     u_{B}(a) = \sum_{b \in \Gamma(a,\optcpl)} d(a,b) \pi(a,b).
\end{align}

\end{definition}

We need to alter the definition of group bias to account for the fact that we can only compare an individual $a \in G \subset A$ to their counterparts in $B$.

\begin{definition}[Group Bias]
The bias (comparatively) experienced across policies for a group $G$ within population $A$ when compared to population $B$ is measured by:
\begin{align}
    U_B(A \cap G) = \sum_{g \in A \cap G} u_{\mathcal{B}}(g).
\end{align}
\end{definition}

Finally, if we decompose the datasets $A$ and $B$ into disjoint groups $A = \{A_1, A_2, \dots A_n \}$ and $B = \{ B_1, B_2, \dots, B_m \}$, then the group bias measurement in population $\A$ relative to population $\B$ can be additively decomposed as
\begin{align*}
    U_B(A_i) = \sum_{j \leq m} U_{B_j}(A_i)
\end{align*}
where $U_{B_j}(A_i)$ is the amount of bias measured in group $A_{i}$ within population $\A$ that is the result of comparison with individuals in group $B_{j}$ within population $\B$. We make use of this framework to compare how individuals in the COMPAS dataset are mapped across race and criminal history-based groups in Section~\ref{sec:compas} and how individuals in the German credit data are mapped across age and sex-based groups in Section~\ref{sec:recourse}. 

In the case that an individual $a$ in group $A$ is very different from everyone in group $B$, the individual bias in Definition~\ref{individual_bias} will be high regardless of who $a$ is mapped to. Thus, to detect such outliers, we can normalize the individual bias by the individual's own worst-case-scenario, i.e. when they have all of their mass mapped to the individual who is farthest from them:
 \begin{equation}
     u^*_B(a) = \frac{n_1 \sum_{j=1}^{n_2} \optcpl(a, b_j) d(a, b_j)}{\max_{j = 1:n_2} d(a, b_j)}.
 \end{equation}
This new normalized individual bias can then be used to differentiate between cases when an individual experiences extreme bias or merely has no equivalent comparison in feature space. 

\subsection*{COMPAS dataset}

COMPAS is a commercial tool used to predict recidivism risk of individuals awaiting trial. The COMPAS dataset contains data compiled by Propublica \cite{larson2016we} on 6172 people who were scored by the COMPAS algorithm in Broward County, Florida from 2013-2014. Each person has a predicted recidivism score as well as a binary flag to indicate whether or not they recidivated. There are 3,175 Black defendants, 2,103 white defendants, and 2,809 defendants who recidivated within two years in this sample. It contains 9 features: two year recidivism, number of priors, age above 45, age below 25, female, misdemeanor, ethnicity, and predicted recidivism probability. We one-hot encode the categorical variables and normalize the features prior to fitting our models.

\subsection*{German credit dataset}

The German credit dataset \cite{German} contains $1000$ instances with $26$ features each of loan applications at a bank. It includes applicant profiles as well as their classification as a bad ($n_1 = 300$) or good ($n_2 = 700$) credit risk.  We one-hot encode categorical variables and then z-score normalize all features before fitting our logistic regression model.

Table~\ref{tab:german_credit} describes how we classified features as actionable, non-actionable, and immutable as inspired by \cite{chen2020strategic}.

\begin{table}[t]
    \centering
    \begin{tabular}{lll}
        \hline
        Variable  & Group \\
        \hline
        Savings & Actionable \\
        Credit history & Actionable & \\
        Other installments & Actionable \\ 
        Co-applicant or guarantor & Actionable \\ 
        Number liable individuals &  Actionable \\ 
        Unemployed  & Actionable  \\
        Property owned & Actionable&  \\
        Number existing credits & Actionable\\
        Installment rate & Actionable\\
        Credit amount & Non-actionable  \\
        Credit duration & Non-actionable\\ 
        Credit purpose & Non-actionable  \\
        Telephone  & Non-actionable \\ 
        Sex & Immutable \\
         Age & Immuatble \\
        Foreign national & Immutable \\
        Years at job & Immutable \\
        Skilled employee & Immutalbe \\
         Permanent resident since  & Immutable \\ 
         \hline 
    \end{tabular}
    \caption{Actionable, non-actionable, and immutable features in the German credit dataset.}
    \label{tab:german_credit}
\end{table}

\begin{figure}[H]
    \centering
    \includegraphics[width = 0.5\textwidth]{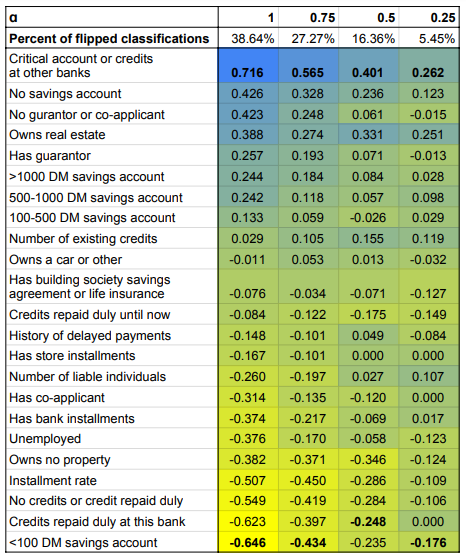}
    \caption{Average changes in actionable features in the German credit dataset after individuals with bad credit were interpolated (at level $\alpha$) to their counterparts with good credit. }
    \label{fig:german_changes}
\end{figure}

For each individual labeled as having bad credit, we interpolate their actionable features with the actionable features of those they were mapped to (Equation~\ref{eq:interp}). Then, we reclassify these individuals. In Figure~\ref{fig:german_changes}, we present the percentage of individuals who were reclassified as having good credit and record the average change for each actionable feature for varying values of $\alpha$.

\end{document}